# Verification Without Distrust: Reframing User-Side Oversight as Routine Epistemic Governance in Everyday Human–Chatbot Interaction

Aung Pyae
*International School of Engineering,*
*Faculty of Engineering*
*Chulalongkorn University, Thailand*
aung.p@chula.ac.th

***Abstract*—Research on human–AI interaction has long framed verification of system outputs as a trust-contingent behavior that better-calibrated trust should reduce. We test this assumption in everyday human–chatbot interaction through a mixed-methods survey of 153 frequent chatbot users. Contrary to the canonical prediction, we find no detectable association between trust and verification, with the result robust across sensitivity analyses. Three further user-side practices — refinement, correction, and approval before automated actions — are widely endorsed and positively associated with satisfaction. The data reveal a substantive distinction between evaluative oversight (trust-decoupled, weakly tied to satisfaction) and interventionist oversight (weakly trust-correlated, strongly tied to satisfaction). A medium-to-large satisfaction–control gap shows that effective task outcomes do not produce a felt sense of agency. Qualitative findings identify instrumental mental models, failure-mode-specific doubt, and demand for epistemic infrastructure. We reframe user-side oversight as routine epistemic governance compatible with trust, and derive four design directions for scaffolded oversight in conversational AI.**

***Keywords—human–chatbot interaction; human–AI interaction; trust calibration; user-side human-in-the-loop; verification; epistemic governance; user agency***

## I. Introduction

AI-powered chatbots—particularly those built on large language models (LLMs)—are now embedded in everyday study, work, and creative practice [1]. As they have scaled, so has research on the human side of interaction: how users form expectations, calibrate reliance, and manage the uncertainty of probabilistic outputs [2], [3]. Within this literature, trust is a central construct, defined as an attitude — the belief that a system will help achieve one's goals under uncertainty and vulnerability [4]. The dominant logic, inherited from work on trust in automation, treats trust and verification as endpoints of a single continuum: greater trust implies less verification, and design that builds appropriate trust should, in turn, reduce the need for users to check outputs [4], [5], [8].

This carries an implicit prescription: if we can build chatbots that users trust appropriately, oversight should become unnecessary; transparency interventions are typically motivated this way [6], [7]. Yet the assumption rests on an empirical claim that has received surprisingly little direct testing in everyday human–chatbot interaction: does trust actually reduce verification when the interaction is open-ended, iterative, and not tied to a binary accept/reject decision?

This paper presents evidence that it does not—at least among frequent users. In a mixed-methods survey of 153 frequent chatbot users, we find no detectable association between trust and verification ($r = .01$, 95% CI [−.15, +.17], $p = .90$), with the confidence interval ruling out moderate inverse linear relationships. Beyond verification, three additional user-side practices—refinement, correction, and oversight of automated actions—are likewise widely endorsed (67–75% top-2) and positively associated with satisfaction. We term these collectively user-side human-in-the-loop (HITL) practices: the spontaneous, self-directed work through which everyday users govern chatbot outputs without dedicated system support. The HITL concept has been studied primarily as a system property [9], with emerging work involving end-users in fairness-related feedback loops [10]; the broader user-enacted side remains empirically underspecified.

Three gaps motivate this study. First, the trust–verification relationship in everyday chatbot use has not been directly tested; prior work has examined it primarily in structured decision support [5], [12]. Second, the prevalence and structure of user-side HITL practices, and their relationship to satisfaction and felt agency, are empirically underspecified. Third, design support for user-enacted oversight is insufficiently articulated [6], [11]: prior work emphasizes features that build trust to reduce checking — with notable exceptions such as Amershi et al.'s guidelines [13], which touch on user-feedback scaffolding without making it the central design target — rather than infrastructure that scaffolds the checking users already perform. We address these gaps through three research questions. RQ1: How prevalent are user-side oversight practices among frequent chatbot users, and how do they relate to trust? RQ2: How do these practices relate to satisfaction and perceived control? RQ3: What design features do users identify as supportive of their oversight practices?

This work makes five contributions: (1) empirical evidence that among frequent users verification shows no detectable association with trust, complicating the reliance-calibration paradigm; (2) a distinction between evaluative oversight (verification: no detectable trust association, weakly tied to satisfaction) and interventionist oversight (refinement, correction, approval: weakly trust-correlated, strongly tied to satisfaction); (3) a satisfaction–control gap ($d = 0.72$) showing that effective task outcomes do not entail felt agency; (4) a theoretical synthesis positioning user-side human-in-the-loop as a fast-process complement to system-side HITL, organized around the three findings above (Fig. 1); and (5) four design directions for scaffolded oversight — provenance cues, process transparency, correction persistence, and governable preference memory — grounded in user-reported needs.

## II. Related Work

### A. Trust, Reliance, and the Calibration Paradigm

Lee and See [4] defined trust as the attitude that an agent will help achieve a goal under uncertainty, and articulated appropriate reliance as the design objective. The framework treats reliance — the decision to use system output — as inversely proportional to trust. The implicit assumption, carried into subsequent literature on user oversight, is that verification (checking that output before acting on it) will track this same inverse pattern, since both reflect underlying uncertainty about reliability. Hoff and Bashir's review [8] subsequently synthesized the empirical literature into a three-layered model (dispositional, situational, learned), formalizing trust as a multidimensional construct. Within this paradigm, recent work has shown that the trust–reliance relationship is more complex than the inverse model suggests. Bansal et al. found AI explanations did not uniformly improve human–AI team performance and could increase reliance on incorrect recommendations [5]. Lai et al.'s survey documented that confidence displays, explanations, and accuracy feedback shape reliance in often unexpected ways [12]. Bussone et al. observed in clinical decision support that fuller explanations could simultaneously increase trust and produce over-reliance [3]. Yet none of this work directly tests whether trust reduces verification in the iterative, open-ended setting of everyday chatbot use.

A further complication concerns the direction of reliance bias itself. Empirical work has shown that users sometimes display algorithm appreciation, weighting algorithmic advice more heavily than human advice in numerical-prediction tasks [20], while in other contexts they display algorithm aversion, abandoning algorithmic recommendations after observing failure even when the algorithm outperforms human alternatives [21]. Both patterns are difficult to reconcile with a simple inverse trust–verification continuum, since they suggest reliance is shaped by the salience of recent errors and contextual cues rather than by stable trust attitudes. Two structural features distinguish everyday chatbot interaction from the decision-support contexts in which this literature was developed. First, the unit of decision differs: in classical reliance studies a user accepts or overrides a single recommendation, whereas in chatbot interaction a user iteratively reformulates queries, refines outputs, and recombines partial responses across exchanges. Second, the cost of verification differs: in many decision-support settings, independent verification is effectively impossible at decision time, whereas in chatbot interaction users can quickly cross-check facts against external sources, regenerate responses, or ask the system to explain its reasoning. These differences leave open the possibility that, in conversational settings, verification operates as a routine practice that is decoupled from — rather than calibrated against — trust.

### B. Human–Chatbot Interaction and User-Enacted Practice

Conversational agents have evolved alongside advances in natural language processing [1]. The literature has identified system-level factors shaping user experience: explainable interface design [6] and customer-facing compliance dynamics [14] — mostly within a system-centric logic in which design features drive user perceptions. Luger and Sellen documented a persistent gulf between user expectations and conversational-agent behavior, with users developing workarounds when system behavior is unpredictable [15]. Suchman's earlier work established the principle: plans embedded in system design are never sufficient to account for the contingencies of actual use [16]. LLM-based chatbots intensify these dynamics; recent work has begun documenting user practices for probing, iterating, and constraining system behavior — for example, Wu et al.'s study of prompt chaining [17] — but a broader empirical characterization of these practices' prevalence, structure, and relationship to outcomes in human–chatbot interaction remains limited.

### C. Human-in-the-Loop and Meaningful Control

HITL has multiple meanings. In machine learning, it denotes architectures incorporating human feedback into training, with active development in domains such as medical AI [9]. In HCI, it has been reframed as an interaction-level concept emphasizing meaningful human control. Shneiderman's human-centered AI framework calls for systems that are reliable, safe, and trustworthy, with human control embedded throughout interaction rather than delegated solely to back-end automation [11]. Amershi et al.'s eighteen guidelines operationalize this principle, foregrounding feedback, control, and capability communication [13]. Wu et al. demonstrated that prompt chaining can increase the transparency and controllability of LLM interactions [17]. Horvitz's principles of mixed-initiative interaction provided an early articulation of the tension between automated action and user agency [18]. A small but growing body of work has begun engaging end-users directly in HITL processes for specific applications — notably Nakao et al.'s work involving end-users in fairness-related feedback for ML systems [10]. These efforts remain confined to specialized contexts (algorithmic fairness, model debugging) and do not address the routine, everyday epistemic practices through which users govern probabilistic chatbot outputs. This body of work emphasizes system-implemented HITL—mechanisms designed by developers; the complementary user-side of everyday chatbot use has received less attention.

## III. Method

### A. Design and Participants

We conducted a cross-sectional mixed-methods online survey (quantitative scaled items + open-ended prompts), recruiting via convenience sampling from technology-oriented online communities. Eligibility required prior chatbot experience and adult age; the survey opened with screening questions on age and chatbot-use frequency, and participants who selected "Under 18" or "Never" were retained in the raw dataset but flagged as ineligible and excluded from primary analyses through the sensitivity check described below. Participation was voluntary, anonymous, and uncompensated; informed consent was obtained upfront and no personally identifiable information was collected, in accordance with institutional ethical guidelines.

The final sample (N = 153) was predominantly young (56.2% aged 18–24; 24.8% aged 25–34), student-enrolled (69.9%), and STEM-oriented (Computing & IT 47.1%; Engineering 28.8%), with substantial frequency of chatbot use (67.3% daily; 90.8% at least weekly); self-rated proficiency was Intermediate (52.3%), Beginner (26.8%), Advanced (17.6%), Novice (0.7%), and Expert (0.7%); 2.0% did not respond. This sample — frequent chatbot users with mixed self-rated proficiency, drawn predominantly from STEM disciplines — is well-suited for detecting whether oversight practices have stabilized into habit among users with substantial exposure to chatbots. We acknowledge that generalization to non-STEM, older, institutionally regulated, or low-exposure populations requires replication; we develop this boundary condition in §V. Seven participants fell outside eligibility criteria (five under 18; two non-users); sensitivity analyses on the compliant subsample (N = 146) confirmed all findings remain robust.

### B. Instrument and Constructs

The instrument comprises seven sections (demographics, mental models, trust, oversight, perceived control, satisfaction, design preferences) on a 5-point Likert scale. Trust was operationalized as a generalized dispositional orientation following Lee and See [4] and Hoff and Bashir [8] ("I generally trust the chatbot's answers"). Four user-side oversight behaviors map onto practices in [13] and [4]: verification ("Before I act on chatbot information, I double-check it"), refinement ("I often ask the chatbot to rephrase, shorten, or improve its answer"), correction ("I feel comfortable telling the chatbot when it's wrong"), and oversight ("I prefer to confirm or approve the chatbot's actions before it does anything automatically"). The mental-model item draws on [15]; perceived control captures the felt-agency dimension of meaningful human control [11]; satisfaction follows standard task-experience operationalization. Distributions span the full scale with meaningful dispersion.

Single items rather than multi-item validated scales reflect a deliberate trade-off given the brief, uncompensated format. They capture narrow behavioral tendencies adequately but cannot capture broad latent constructs such as the three-layered trust hierarchy [8] with full psychometric precision. Because measurement noise tends to attenuate rather than inflate associations, the trust–verification null is best read as the absence of a moderate-or-larger inverse relationship rather than proven independence. We mitigate this by treating items as behavioral frequency indicators, triangulating with open-ended responses, and reporting Pearson and Spearman correlations in parallel. Replication with multi-item validated instruments [8] is a priority for future work.

### C. Analysis

Quantitative analysis includes descriptive statistics, Pearson and Spearman correlations with Fisher-transformed 95% CIs, and moderated regression with mean-centered predictors. The satisfaction composite (G1+G2) shows excellent internal consistency ($\alpha = .90$). The four-item oversight engagement score ($\alpha = .68$) is used only for the moderation test; individual-behavior analyses are prioritized. VIFs are all $\leqslant 1.34$ (no multicollinearity). At N = 153, the study has 80% power to detect $r \geqslant .23$ ($\alpha = .05$, two-tailed). Open-ended responses were analyzed via thematic analysis following Braun and Clarke's six phases [19]. Analysis was single-coder; we acknowledge this as a limitation and treat qualitative findings as convergent evidence rather than a standalone contribution. As a quality-screening sensitivity check, five participants with identical Likert responses across all ten items were flagged and excluded (screened N = 148). The trust–verification null is preserved ($r = -.12$, 95% CI [−.28, +.04], $p = .14$; $\rho = -.15$, $p = .06$), the satisfaction–control gap persists ($M_{diff} = 0.91$, $t(147) = 8.96$, $p < .001$, $d = 0.74$), and all positive associations between interventionist oversight behaviors and satisfaction remain (Refinement and Approval both $p < .001$; Correction $p = .001$). No findings are artifacts of low-engagement responding.

## IV. Results and Findings

### A. Prevalence of Oversight (RQ1)

Table I summarizes descriptive statistics. All four oversight behaviors were widely endorsed: refinement (M = 4.05, 75.2% top-2), correction (M = 4.07, 73.9%), verification (M = 3.83, 67.3%), and oversight of automated actions (M = 3.84, 67.3%). Roughly one in ten or fewer disagreed with any individual oversight item (verification 10.5%, refinement 5.2%, correction 9.2%, approval 11.1%). Among frequent users, active governance of chatbot outputs is the norm rather than an exceptional response to system failure. In contrast, generalized trust was moderate (M = 3.20, SD = 0.85), with the modal response at the scale midpoint (47.1% selected 3). Just over one-third (36.6%) agreed they trust chatbot outputs; 16.3% disagreed.

TABLE I. DESCRIPTIVE STATISTICS FOR SCALED VARIABLES (N = 153)

| Item | Construct | M | SD | Mdn | %4-5 |
|---|---|---|---|---|---|
| **B2** | **Tool–teammate scale** | 3.12 | 1.19 | 3 | 39.2% |
| **C1** | **Trust** | 3.20 | 0.85 | 3 | 36.6% |
| **C2** | **Credibility cues** | 3.60 | 1.09 | 4 | 56.2% |
| **D1** | **Verification** | 3.83 | 1.00 | 4 | 67.3% |
| **E1** | **Refinement** | 4.05 | 0.94 | 4 | 75.2% |
| **E2** | **Correction** | 4.07 | 1.13 | 4 | 73.9% |
| **F2** | **Oversight (approval)** | 3.84 | 1.11 | 4 | 67.3% |
| **F1** | **Perceived control** | 3.33 | 1.19 | 3 | 45.1% |
| **G1** | **Task facilitation** | 4.26 | 0.94 | 5 | 83.0% |
| **G2** | **Enjoyment** | 4.17 | 0.92 | 4 | 80.4% |

### B. No Detectable Trust–Verification Association (RQ1)

The central test of this study examines whether trust and verification are inversely related. We find no evidence that they are. The Pearson correlation was $r = .01$, 95% CI [−.15, +.17], $p = .90$. Spearman's rank correlation confirmed the pattern: $\rho = -.08$, $p = .33$. The 95% CI rules out moderate or larger negative linear relationships in the population, although small effects within the confidence interval ($|r| \leq .17$) and nonlinear or subgroup-specific patterns cannot be excluded by this analysis. The result is not attributable to restricted variance (both items range across the full 1–5 scale with meaningful dispersion) nor to insufficient power (the study had 80% power to detect $r \geq .23$). The eligibility-compliant subsample (N = 146) reproduced the result ($r = -.03$, $p = .70$). We therefore frame the result as no detectable association rather than as proven independence: a strong null on the canonical inverse relationship, but compatible with subtler structure that future work with multi-item scales [8] and behavioral measures should test. The null is specific to verification: refinement, correction, and oversight showed small positive Pearson correlations with trust ($r = .19$–$.23$, 95% CIs all excluding zero) but did not reach significance under Spearman analysis ($\rho = .12$–$.14$).

### C. Evaluative vs. Interventionist Oversight (RQ2)

Both trust and oversight engagement were associated with satisfaction (Trust–Sat r = .50, 95% CI [.38, .61]; Oversight composite–Sat r = .51, 95% CI [.38, .61]; both $p < .001$). Table II reports individual-behavior correlations. A clear pattern emerges: refinement showed the strongest satisfaction association (r = .48), followed by oversight (r = .39), correction (r = .38), and verification (r = .21, with non-significant Spearman ρ = .10). The four behaviors fall into two structurally distinct clusters. Verification is evaluative—users check outputs without changing them—showing no detectable trust association and only weakly tied to satisfaction. Refinement, correction, and approval are interventionist—users actively shape outputs—and are weakly trust-correlated and strongly tied to satisfaction. Treating user-side HITL as a single composite obscures this; the two clusters relate differently to trust and to satisfaction, and we propose future work treat them as theoretically distinct.

A moderated regression (Sat ~ Oversight + Trust + Oversight × Trust, mean-centered) explained 44% of variance, $F(3,149) = 39.27$, $p < .001$ (all VIFs ⩽ 1.34). Both oversight (b = 0.385, $p < .001$) and trust (b = 0.407, $p < .001$) were independent predictors. The interaction was negative and significant in the full sample (b = −0.165, p = .011), suggesting oversight contributes more to satisfaction when trust is lower; we treat this as tentative because the eligibility-compliant subsample (N = 146) attenuated it to marginality (b = −0.140, p = .07). The model also likely overstates explained variance because predictors and outcome share self-report common-method variance.

TABLE II. INDIVIDUAL OVERSIGHT BEHAVIORS AND OUTCOMES (N = 153)

| Behavior | Sat (r [CI] / ρ) | Ctrl (r/ρ) | Trust (r/ρ) |
|---|---|---|---|
| **Verification** | .21** [.05,.36] / .10 | .06 / .01 | .01 / −.08 |
| **Refinement** | .48*** [.34,.59] / .36*** | .22** / .15 | .23** / .13 |
| **Correction** | .38*** [.23,.51] / .35*** | .30*** / .28*** | .22** / .12 |
| **Oversight** | .39*** [.25,.52] / .33*** | .19* / .22** | .19* / .14 |

*Each outcome cell shows Pearson r / Spearman ρ. 95% CI shown for r(Sat). Sat = satisfaction composite (G1+G2); Ctrl = perceived control (F1); Trust = generalized trust (C1). *p<.05; **p<.01; ***p<.001; unmarked = ns.*

### D. The Satisfaction–Control Gap (RQ2)

Users reported high satisfaction (M = 4.22; 78.4% scoring ≥ 4 on the composite, with 83.0% endorsing G1 and 80.4% endorsing G2 individually) but substantially lower perceived control (M = 3.33, 45.1% top-2). The paired difference was robust (M_diff = 0.88, 95% CI [0.69, 1.08], $t(152) = 8.88$, $p < .001$, Cohen's d = 0.72). The satisfaction–control correlation was significant but only moderate (r = .32, 95% CI [.17, .46], $p < .001$; ρ = .25, p = .002), confirming these are partially dissociable experiential dimensions. A regression of perceived control on the oversight composite, trust, and their interaction (paralleling the satisfaction model) explained only 14% of variance, $F(3,149) = 7.96$, $p < .001$, indicating that felt agency depends on factors beyond these predictors. The gap is a substantive finding: effective task outcomes do not entail felt control, with direct consequences for how human–chatbot interaction should be evaluated and designed.

### E. What Gets Verified

On a multi-select item ("What kinds of outputs do you always verify?") participants most commonly reported checking factual content (68.0%), numbers (51.6%), and work-related output (47.7%). Creative ideas were verified by 36.6%, and only 8.5% reported rarely verifying. The breadth—spanning facts, numbers, and work content—is consistent with framing verification as routine epistemic hygiene rather than failure-mode-triggered checking.

### F. Qualitative Findings

Among 132 substantive responses describing the chatbot's role, the dominant framing was instrumental — chatbots as tools, assistants, or efficiency aids ("a tool to get an answer quickly but still need to recheck the accuracy"; "my secretary who knows everything"). Roughly 15–20% used collaborative or relational language (mentor, teammate, friend, tutor, advisor), but relational framings frequently coexisted with acknowledgment of oversight needs ("Teammate without judgment but you have to double check with it"; "a decent but dumb and silly machine"). In 108 trust-or-doubt accounts, trust was conditional and failure-mode-specific. The most frequently cited triggers were mathematical errors and fabricated references ("Sometime AI chatbot had made mistakes like Calculus or Discrete or Mathematics so I always recheck final results or step by step calculations"; "it tends to make up stuff and references"), alongside confidently-wrong answers ("Which is the bigger number? 9.11 or 9.9. The chatbot answered… wrong with full confidence") and sycophancy ("it's a bit suspicious when the chatbot starts apologizing when I complain and follows my way without proper logical reasoning"). Verification was repeatedly described as a standing practice rather than a response to distrust: "I usually trust chatbot response when I provide all necessary data to it. However, i [sic] always check the response." This narrative framing maps directly onto the quantitative null.

When asked what would increase confidence without requiring double-checking (n = 77), the most frequent request was source attribution ("verified sources or expert references"), followed by reasoning transparency. A minority view captured the underlying epistemic position: "Nothing. This is an inherently probabilistic tool." On learning from corrections (n = 68), participants articulated two desires — persistent preference memory and correction persistence ("if I correct them on something, [they should] consider it the next times and not make the same mistake") — though some noted privacy or epistemic concerns about uncontrolled adaptation ("there is a risk of misinformation if chatbot learns directly from users"), suggesting preference memory must be governable: inspectable, configurable, and reversible.

## V. DISCUSSION

### A. Verification as Routine Epistemic Governance

The central finding — that trust and verification do not move together in everyday chatbot use — has theoretical force beyond its statistical form. The reliance-calibration paradigm was developed for discrete decisions made under time pressure with limited intermediate steps [4], [5]. Everyday human–chatbot interaction is structurally different: iterative, open-ended, and conducted with outputs that are explicitly probabilistic. In that setting, verification is better understood as routine epistemic governance — the ongoing maintenance of well-grounded belief through habitual checking, analogous to proofreading or double-checking calculations in skilled professional work — than as a trust-contingent response. The qualitative evidence supports this reading: participants reporting moderate trust described verification as routine in the same breath ("I usually trust… However, I always check"), a co-occurrence more naturally described as a dual-process pattern — trust as dispositional orientation, verification as enacted governance — than as endpoints of a single continuum. Two boundary conditions matter. First, the null persists across the proficiency and disciplinary subgroups we examined, including self-rated beginners and non-STEM participants; generalization to novices, non-technical, or institutionally regulated populations nonetheless requires replication, and the pattern may be strongest where epistemic governance has become habitual. Second, our null applies to moderate or larger inverse linear relationships; small or nonlinear effects are not excluded, and future work with multi-item validated trust instruments [8] and behavioral measures should probe more subtle structure. The contribution is a principled push-back: the assumption that better-calibrated trust will reduce checking needs much more empirical support than it has received in conversational settings. Fig. 1 summarizes the position — trust and verification as parallel processes rather than continuum endpoints, user-side oversight as heterogeneous, and effective task outcomes as separable from felt agency.

### B. *Evaluative vs. Interventionist Oversight*

A second contribution concerns the internal structure of user-side HITL. Treating oversight as a single category obscures a distinction between evaluative practices (verification — checking outputs without changing them) and interventionist practices (refinement, correction, approval — actively shaping outputs). Active iteration carried the strongest satisfaction premium, suggesting that satisfaction in human–chatbot interaction is produced through dialogue rather than delivered intact in a single response. Design implications diverge: evaluative oversight calls for infrastructure that lowers the cost of checking (provenance, source attribution, reasoning traces); interventionist oversight calls for infrastructure that makes intervention legible and consequential (correction persistence, refinement scaffolding, governable preference memory). Rather than treating oversight and trust as competing, the present pattern suggests they are compatible, with engagement mattering most under lower trust. Trust need not be reduced, but paired with infrastructure that supports users' existing oversight practices.

### C. *The Satisfaction–Control Gap as Parallel Finding*

A third contribution concerns the dissociation between satisfaction and felt agency. The medium-to-large gap (d = 0.72) shows that users extract value from current chatbots while still experiencing the exchange as partially opaque and difficult to steer — a pattern previously described qualitatively in conversational agents [15], here quantified and shown to be only weakly explained by trust or oversight behavior. Whatever determines felt control lies largely outside the constructs measured here: the levers users have for checking, refining, and correcting do not, in their current form, translate into a sense of being in charge. Meaningful human control, in Shneiderman's sense [11], requires that interventions are legible (visible as they happen) and consequential (producing predictable, persistent change). The gap suggests that consequentiality is what is currently missing, even when satisfaction is high. Evaluation regimes should therefore track perceived control alongside satisfaction — otherwise designs that succeed on satisfaction may quietly fail on agency — and design effort should target the consequentiality of intervention (whether corrections persist across turns and sessions, whether refinements produce predictable changes) as a first-class outcome rather than a side effect of usability.

### D. *Implications for a Theory of Human–Chatbot Interaction*

Taken together, the three findings sketch a theoretical position that complements rather than displaces the reliance-calibration paradigm. The paradigm assumed a single user-side construct, trust, organizing a single user-side response, reliance. Our results suggest everyday human–chatbot interaction is better characterized by two coupled processes on different timescales. The first is slow and dispositional: trust reflects an accumulated attitude that shifts only as failure modes accumulate over extended use. The second is fast and enacted: verification, refinement, correction, and approval are situated practices that govern outputs at the moment of use. The fast process is sustained by interface affordances and habit rather than moment-to-moment trust judgments — which is why verification persists as a widespread practice even among participants reporting relatively high trust. Within the fast process, our data support a distinction between evaluative and interventionist oversight. Evaluative oversight (verification) is informational: it produces knowledge about whether an output is acceptable but does not change it. Interventionist oversight (refinement, correction, approval) is generative: it shapes the output toward user goals. The two differ in their relations to satisfaction (weak vs. strong) and to trust (no detectable association vs. weakly positive). Theories of human–AI interaction grounded in single-shot decisions [4], [5], [12] tend to collapse this distinction; conversational settings make it visible.

The satisfaction–control gap (d = 0.72) adds a third element: even when the fast process is engaged effectively, users do not necessarily experience the interaction as one they are steering. Satisfaction and meaningful human control [11] are partially dissociable in current chatbot interfaces, and perceived control depends on a property our instruments do not measure directly — the consequentiality of intervention, whether oversight actions visibly shape future system behavior. The qualitative demand for correction persistence and inspectable preference memory points to consequentiality as the missing variable. We therefore propose user-side HITL as a construct complementary to system-side HITL studied in machine-learning literatures [9], with limited prior engagement of end-users in specialized contexts such as algorithmic fairness [10]: situated, fast-process practices through which users govern probabilistic outputs, jointly determined by interface affordances, habit, and the consequentiality of intervention rather than by trust alone.

**User-Side Oversight in Human–Chatbot Interaction**

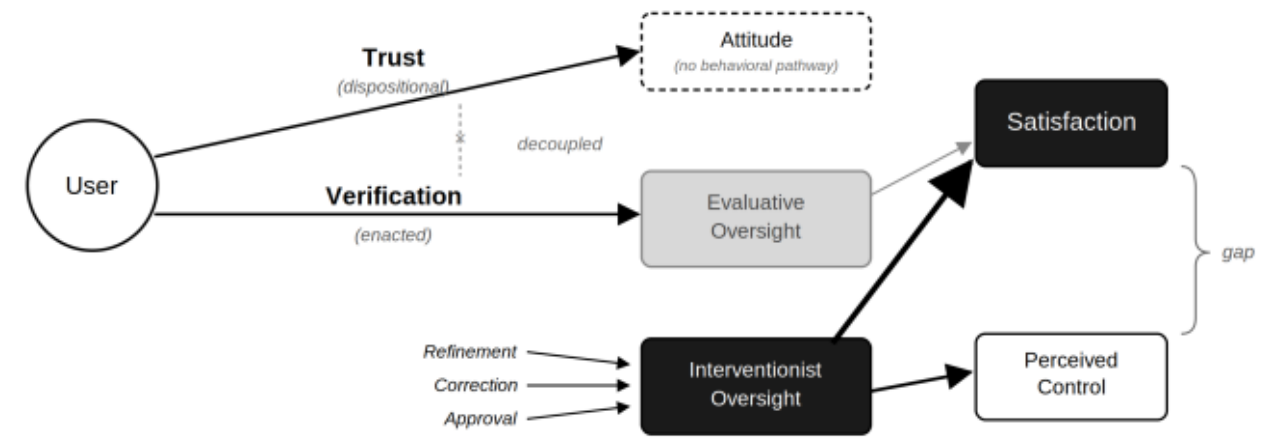

**Fig. 1.** Conceptual model of user oversight in human–chatbot interaction. Trust (dispositional) and verification (enacted) function as parallel decoupled processes; evaluative oversight (verification) relates only weakly to satisfaction, while interventionist oversight (refinement, correction, approval) is tied to both satisfaction and perceived control. The satisfaction–control gap motivates scaffolding oversight as a first-class interaction mode.

## E. Design Directions for Scaffolded Oversight (RQ3)

Four design directions follow, each grounded in user-reported needs. (i) Provenance and source attribution. The most frequent qualitative request was for embedded sources and citations; provenance should be treated as workflow infrastructure that lowers the cost of evaluation [6], [7], surfaced inline for factual and numerical content. (ii) Process transparency. Reasoning traces and uncertainty indicators support users in evaluating not only what the chatbot produced but how it arrived there, with chained or step-decomposed outputs as in [17]. (iii) Correction persistence. Corrections should propagate within and, where appropriate, across sessions; observable change — showing users their corrections shaped subsequent behavior — may matter more for perceived control than the corrections themselves. (iv) Governable preference memory. Personalization should be inspectable, configurable, and reversible, in the spirit of mixed-initiative interaction [18]. These are testable design hypotheses sharing one orientation: rather than eliminating user oversight, conversational AI should scaffold oversight as a first-class interaction mode.

## F. Limitations and Future Work

Five limitations qualify these inferences. Single-item measurement frames the trust–verification null as ruling out a moderate-or-larger inverse relationship rather than establishing independence. The convenience sample skews young, student-enrolled, and STEM-oriented, calling for replication in non-technical, older, and institutionally regulated populations. Self-report invites social-desirability bias; behavioral records would mitigate this. Trust was measured as a generalized disposition; content-specific scales may surface obscured relationships. Shared-method effects plausibly inflate the satisfaction model. Future work should replicate the null with multi-item scales and behavioral checking records, experimentally test the four design directions, and extend to lower-frequency and institutionally regulated populations — though present subgroup evidence suggests the result is unlikely to be confined to STEM-trained or advanced users.

# VI. Conclusion

This paper investigated user-side oversight in everyday human–chatbot interaction and the assumed inverse trust–verification relationship. Among frequent users, trust and verification show no detectable association, contrary to the reliance-calibration paradigm. Oversight is structurally heterogeneous — evaluative practices (verification) are trust-decoupled and weakly tied to satisfaction; interventionist practices (refinement, correction, approval) are weakly trust-correlated and strongly tied to satisfaction — and a satisfaction–control gap shows that effective task outcomes do not entail felt agency. Qualitative evidence converges on a consistent design need: epistemic infrastructure that makes oversight more efficient rather than unnecessary. We reframe user-side oversight as routine epistemic governance compatible with trust, and shift the design question from "how do we build trust so users stop checking?" to "how do we support the checking users already do?"